# CaseCohortCoxSurvival: an R Package for Case-Cohort Inference for Relative Hazard and Pure Risk under the Cox Model

Lola Etiévant[1], Mitchell H. Gail[1]

lola.etievant@nih.gov          ORCID 0000-0001-7562-3550

gailm@mail.nih.gov             ORCID 0000-0002-3919-3263

[1] National Cancer Institute, Division of Cancer Epidemiology and Genetics, Biostatistics Branch, 9609 Medical Center Drive, Rockville, MD 20850-9780

# Table of contents

# Abstract

The case-cohort design allows analysis of multiple endpoints and only requires covariates to be measured for cases and non-cases in a random subcohort from the cohort. Stratification of subcohort sampling and weight calibration increase efficiency of estimates of log-relative hazards and covariate-specific pure risk, but they may require specifically adapted variance estimators. Some recent articles in epidemiology and medical journals used an inappropriate "robust" variance estimator. In addition, stratification, weight calibration and analysis of pure risk seem underutilized in case-cohort studies, possibly because practitioners are put off by the varied technical methodologic literature and lack of convenient software. We recently proposed a unified approach to variance estimation for Cox model log-relative hazards and pure risks, and we implemented it in an R package, [CaseCohortCoxSurvival](), available on CRAN, that allows appropriate and convenient analysis of case-cohort data, with and without stratification, weight calibration, or missing at random phase-two data. Here we illustrate how easy it is to use [CaseCohortCoxSurvival]() to analyze data from the Prostate, Lung, Colorectal and Ovarian Cancer Screening Trial to estimate pure covariate-specific risk of prostate cancer incidence with various case-cohort design and analysis options. These analyses also indicate situations where the simple "robust" variance is too large.

Cohort studies of covariate-specific risk of a rare endpoint such as a cancer require large samples, and it may be too expensive to measure covariates such as biomarkers on all cohort members. Prentice (1) introduced the case-cohort design for estimating log-relative hazards (RH) under the Cox model. Although biospecimens need to be stored on all cohort members, this design reduces costs because covariates such as biomarkers need only be measured for subjects with an event (the cases) and for non-cases in a random sample (the subcohort) from the cohort. In addition, because sampling of the subcohort is performed independently of case status, the same subcohort can be used for different endpoints, such as lung cancer and colorectal cancer.

Subsequent publications described refinements of the original design. Borgan et al. (2) proposed stratifying the subcohort sampling and used design weights for the sampled non-cases, namely the inverse of the stratum-specific sampling fractions. This approach increased efficiency for estimating RH under the Cox model. Calibrating the design weights with information available on the entire cohort was also shown to increase efficiency (3). Although some authors also considered estimation of cumulative baseline hazard and covariate-specific "pure" risk (PR), most of the literature focused on RH.

Analyzing stratified case-cohort data requires a corresponding variance estimator. The "robust variance" estimator, that was proposed by Barlow (4) for estimating RH from unstratified case-cohort data, is too large for estimating RH in stratified designs (5). However, from a review of recent analyses of case-cohort data in medical and epidemiological journals (6), we noted that insufficient attention was paid to specifying the design and that many authors used the "robust variance" with stratified case-cohort data. In addition, weight calibration and analysis of PR were underutilized, and various *ad hoc* methods were used to deal with missing covariate information. We think that the uneven quality of these analyses reflects a highly technical and varied methodologic literature and lack of convenient software adapted to the various design and analysis options. Indeed, while some R packages can estimate the variance of RH from unstratified or stratified case-cohort data (7–9), most do not consider PR and weight calibration. In addition, procedures and code showing how to use survey sampling software to accommodate weight calibration and stratification for RH are no longer online (10,11).

We recently developed a unified approach to variance estimation for the Cox model RH and PR from case-cohort data that relies on linearization and calculating the influence of each case-cohort individual on the parameter estimates (12). The method properly accounts for stratification and weight calibration and can also handle missing at random covariates that are only measured in cases and non-cases in the subcohort. We showed in simulations that these procedures had nominal operating characteristics, including correct coverage of 95% confidence intervals for HR and PR in a variety of settings (12). This work also clarified when the popular robust variance is appropriate.

To overcome impediments to efficient design and analysis of case-cohort studies from lack of convenient software, we implemented the proposed method in the R package [CaseCohortCoxSurvival](). In one line of code, one can provide design parameters such as stratification variables, and analysis parameters such as variables used in the Cox model and variables used for weight calibration. Then the software returns parameter and variance estimates. The goal of this paper is to present the features of CaseCohortCoxSurvival and illustrate its use on data from the Prostate, Lung, Colorectal and Ovarian (PLCO) Cancer Screening Trial (13).

PLCO was a randomized controlled trial of cancer-free individuals to determine the effectiveness of various screening strategies for reducing cancer-specific mortality. We focus on the prostate cancer screening arm of PLCO that included $n = 33{,}547$ individuals aged 49-75 who were recruited in 1993-2001, screened annually for five years for prostate cancer, including prostate-specific antigen (PSA) tests, and followed until 2009 for prostate cancer incidence and until 2015 for mortality. We are interested in predicting the PR of prostate cancer incidence from the following variables that were measured at the time of recruitment (baseline): $X_1$ = indicator of first-degree family history of prostate cancer, $X_2$ = logarithm of continuous prostate specific antigen level in ng/mL, $X_3$ = indicator of being non-Hispanic White, $X_4$ = indicator of being non-Hispanic Black. We assume a Cox proportional hazards model on the age-scale with hazard function $\lambda(t) = \lambda_0(t) \exp(\boldsymbol{\beta}'\boldsymbol{X})$, where $\lambda_0(t)$ is a baseline hazard function which is homogeneous across strata and $\boldsymbol{X} = (X_1, X_2, X_3, X_4)'$. We are interested in computing the PR in age interval $(\tau_1, \tau_2] = (58, 68]$ and for covariate profiles $\boldsymbol{x} = (1, -0.7, 0, 1)'$ and $\boldsymbol{x} = (0, 0.7, 1, 0)'$, where for example $\boldsymbol{x} = (1, -0.7, 0, 1)'$ corresponds to the profile of non-Hispanic Black man with a first-degree family history of prostate cancer and low PSA.

We compare the full cohort analysis with case-cohort designs with or without stratification of the subcohort sampling, with or without calibration of the design weights, and in the presence of missing data and show how CaseCohortCoxSurvival can deal with all these contingencies, summarized in Table 1. In order to define the arguments required by CaseCohortCoxSurvival, we first outline the notation and approach to analysis.

## Notation and Method for Analyzing Case-Cohort Data

In the following sections we define the key elements needed for complete specification and correct analysis of case-cohort data. We give results for stratified sampling of the subcohort. Unstratified sampling corresponds to the special case of one stratum.

**Case-Cohort Design**

Suppose $J$ is the number of strata in the entire cohort (*phase-one* sample), $n^{(j)}$ is the number of subjects in stratum $j$, $j \in \{1, \ldots, J\}$, and $n = \sum_{j=1}^{J} n^{(j)}$ is the number of subjects in the entire cohort. Independently across strata, a fixed number of individuals, $m^{(j)}$, is sampled from stratum $j$ without replacement and independently of case status. The subcohort includes all the sampled subjects from the $J$ strata. The stratified case-cohort or *phase-two* sample consists of all members of the subcohort and any other cases not in the subcohort. We let $\xi_{i,j}$ be the sampling indicator and $w_{i,j}$ be the known design weight of individual $i$ in stratum $j$; $w_{i,j} = n^{(j)}/m^{(j)}$ for a non-case, and $w_{i,j} = 1$ for a case. The special case $J = 1$ corresponds to a non-stratified case-cohort. The covariates in $X$ that are only measured in the phase-two sample are called *phase-two covariates* and denoted $X^{(2)}$. The other covariates in $X$, measured on all cohort members, are called *phase-one covariates* and denoted $X^{(1)}$.

In the PLCO example, we assume that $X_2$ is a phase-two covariate, while $X_1$, $X_3$, $X_4$, $W =$ indicator of PSA > 4 and some other baseline covariates are available on the entire cohort. We base stratification of the subcohort on $W$. We sample $m^{(1)} = 3{,}177$ and $m^{(2)} = 2{,}380$ individuals from the $J = 2$ strata, so that we expect approximately one non-case per case in each stratum; see **Table 2**. We also consider estimation with unstratified case-cohort data. To obtain the unstratified subcohort we sample $m = 4719$ individuals from the entire cohort ($n = 33{,}547$); see **Table 2**. Once these design elements and the Cox model are specified, one can estimate the parameters and use influence function theory to estimate their variance. The following two sections provide formulas for parameter estimation and an overview of two variance estimation methods; for details see (12). Skipping the technical details in these two sections will not impair the ability to use the R software.

**Estimation of log-Relative Hazards, Cumulative Baseline Hazard and Covariate-Specific Pure Risk with Design Weights**

An estimate of the log-relative hazard $\boldsymbol{\beta}$ is obtained by solving the estimating equation

$$U(\boldsymbol{\beta}) = \sum_{j=1}^{J} \sum_{i=1}^{n^{(j)}} \int_t \left\{ X_{i,j} - \frac{S_1(t;\boldsymbol{\beta})}{S_0(t;\boldsymbol{\beta})} \right\} dN_{i,j}(t) = 0, \qquad (1)$$

with

$$S_0(t;\boldsymbol{\beta}) = \sum_{j=1}^{J} \sum_{k=1}^{n^{(j)}} w_{k,j}\, \xi_{k,j}\, Y_{k,j}(t) \exp(\boldsymbol{\beta}' X_{k,j}), \qquad (2)$$

$$S_1(t;\boldsymbol{\beta}) = \sum_{j=1}^{J} \sum_{k=1}^{n^{(j)}} w_{k,j}\, \xi_{k,j}\, Y_{k,j}(t) \exp(\boldsymbol{\beta}' X_{k,j}) X_{k,j}, \qquad (3)$$

and where $Y_{i,j}(t)$ and $dN_{i,j}(t)$ indicate if the individual is at-risk or has the event of interest, respectively, at time/age $t$. We let $\widehat{\boldsymbol{\beta}}$ denote the solution to equation (1) and estimate the baseline hazard point mass at time $t$ non-parametrically (14) by

$$d\widehat{\Lambda}_0(t;\widehat{\boldsymbol{\beta}}) \equiv d\widehat{\Lambda}_0(t) = \sum_{j=1}^{J} \sum_{i=1}^{n^{(j)}} dN_{i,j}(t) / S_0(t,\widehat{\boldsymbol{\beta}}). \qquad (4)$$

The estimated cumulative baseline hazard up to time $t$ is

$$\widehat{\Lambda}_0(t;\widehat{\boldsymbol{\beta}},\hat{\lambda}_0) \equiv \widehat{\Lambda}_0(t) = \int_0^t d\widehat{\Lambda}_0(s), \qquad (5)$$

and the PR for profile $\boldsymbol{x}$ in the time/age interval $(\tau_1, \tau_2]$ is given by

$$\hat{\pi}(\tau_1, \tau_2; \boldsymbol{x}, \widehat{\boldsymbol{\beta}}, d\widehat{\Lambda}_0) \equiv \hat{\pi}(\tau_1, \tau_2; \boldsymbol{x}) = 1 - \exp\left\{ -\int_{\tau_1}^{\tau_2} \exp(\widehat{\boldsymbol{\beta}}'\boldsymbol{x})\, d\widehat{\Lambda}_0(s) \right\}. \qquad (6)$$

**Variance Decomposition and Estimation from Influence Functions**

Barlow (4) proposed the robust variance estimator based on influence functions for Cox model HR with unstratified case-cohort data with design weights. It was later also recommended for PR (10).

We let $\Delta_{i,j}(\widehat{\boldsymbol{\theta}})$ denote the influence of subject $i$ in stratum $j$ on $\widehat{\boldsymbol{\theta}}$, $\widehat{\boldsymbol{\theta}} \in \{\widehat{\boldsymbol{\beta}}, d\widehat{\Lambda}_0(t), \widehat{\Lambda}_0(t), \widehat{\pi}(\tau_1, \tau_2; \boldsymbol{x})\}$. The robust variance estimate of $\widehat{\boldsymbol{\theta}}$ is

$$\widehat{V}_{\text{robust}}(\widehat{\boldsymbol{\theta}}) = \sum_{j=1}^{J} \sum_{i=1}^{n^{(j)}} \Delta_{i,j}(\widehat{\boldsymbol{\theta}}) \Delta_{i,j}(\widehat{\boldsymbol{\theta}})'. \tag{7}$$

Etiévant and Gail (12) instead recommended estimating $\text{var}(\widehat{\boldsymbol{\theta}})$ by

$$\widehat{V}(\widehat{\boldsymbol{\theta}}) = \frac{n}{n-1} \sum_{j=1}^{J} \sum_{i=1}^{n^{(j)}} \frac{\Delta_{i,j}(\widehat{\boldsymbol{\theta}}) \Delta_{i,j}(\widehat{\boldsymbol{\theta}})'}{w_{i,j}} + \sum_{j=1}^{J} \sum_{i=1}^{n^{(j)}} \sum_{k=1}^{n^{(j)}} w_{i,k,j} \sigma_{i,k,j} \Delta_{i,j}(\widehat{\boldsymbol{\theta}}) \Delta_{k,j}(\widehat{\boldsymbol{\theta}})', \tag{8}$$

where $w_{i,k,j} = n^{(j)}/m^{(j)} (n^{(j)} - 1)/(m^{(j)} - 1)$, $\sigma_{i,k,j} = m^{(j)}/n^{(j)} (m^{(j)} - 1)/(n^{(j)} - 1) - (m^{(j)}/n^{(j)})^2$ and $\sigma_{i,i,j} = m^{(j)}/n^{(j)} (1 - m^{(j)}/n^{(j)})$ if individuals $i$ and $k$ in stratum $j$ are both non-cases, $w_{i,k,j} = w_{i,j} \times w_{k,j}$, $\sigma_{i,k,j} = 0$ and $\sigma_{i,j} = 0$ otherwise, $i, k \in \{1, \ldots, n^{(j)}\}$, $k \neq i$, $j \in \{1, \ldots, J\}$. This formula is based on a decomposition of $\text{var}(\widehat{\boldsymbol{\theta}})$ that properly accounts for the subcohort sampling features (12).

In PLCO data, for each sampling design and analysis option we can compare $\widehat{V}_{\text{robust}}(\widehat{\boldsymbol{\theta}})$ with $\widehat{V}(\widehat{\boldsymbol{\theta}})$. We note that $\widehat{V}_{\text{robust}}(\widehat{\boldsymbol{\theta}})$ is valid for estimation with the entire cohort.

**Calibration of the Design Weights**

Calibration of the design weights can improve efficiency of case-cohort studies for Cox model log-RH (3,15) and PR (12). The idea is to use *auxiliary variables* available on the entire cohort and highly correlated with the influences on $\widehat{\boldsymbol{\theta}}$ to perturb the design weights and produce calibrated weights $w^*_{k,j}$ that impart additional information from the auxiliary variables. Parameter estimation with calibrated weights is as in Equations (1), (4), (5) and (6), but with $w^*_{k,j}$ replacing $w_{k,j}$ in Equations (2) and (3). The variance formula differs slightly from Equation (8), because individuals who are in the cohort but not in the phase-two sample influence the calibrated weights and therefore have an influence on $\hat{\boldsymbol{\theta}}$. See Etiévant et al. (12) for the explicit formula.

Auxiliary variables for the log-RH parameters are constructed as in Breslow et al. (3,15). First, phase-two covariates are predicted by regressing them on phase-one covariates, including possibly some covariates not included in the Cox model; weighted regression in the phase-two data is used for this. If a phase-two variable is continuous, then weighted linear regression is used. If a phase-two variable is binary, weighted logistic regression is used. If a phase-two variable is polytomous, then polytomous weighted logistic regression (also called multinomial logistic regression) is used. Then one fits the Cox model with the phase-one covariates $\boldsymbol{X}^{(1)}$ and predicted phase-two covariates $\widetilde{\boldsymbol{X}}^{(2)}$ (that only depend on phase-one data). Letting $\widetilde{\boldsymbol{\beta}}$ denote the estimated log-RH parameters, the influences $\boldsymbol{\Delta}_{i,j}(\widetilde{\boldsymbol{\beta}})$ approximate the true influences and are used as auxiliary variables for log-RH. Auxiliary variables for the PR parameters are constructed as in the approach proposed by Shin et al. (16) for analyzing nested case-control data. First, the case-cohort design weights are calibrated against $\boldsymbol{\Delta}_{i,j}(\widetilde{\boldsymbol{\beta}})$ and a variable that is identically equal to 1. Then one fits the Cox model with $\boldsymbol{X}$ to obtain the estimate $\breve{\boldsymbol{\beta}}$ from the case-cohort data with these calibrated weights. Finally, one multiplies the total follow-up time (on the time interval for which the PR is to be estimated) and the estimated relative hazard $\exp\{\breve{\boldsymbol{\beta}}'(\boldsymbol{X}^{(1)}, \widetilde{\boldsymbol{X}}^{(2)})\}$ to obtain an auxilliary variable for cumulative hazard, and hence for PR. Using both these RH and PR auxiliary variables is the default option ("`Shin`") in CaseCohortCoxSurvival. Using only the RH auxiliary variables is the other option ("`Breslow`" as in references (3,15)). Breslow and Lumley (17) proposed another auxiliary variable for PR, but our unreported simulations indicate that the Shin auxiliary for PR is preferred. In our application to PLCO data, we predict the logarithm of continuous PSA, the phase-two covariate $X_2$, by weighted regression in the phase-two data on baseline variables known on all subjects in the cohort, namely $W$ = indicator of PSA > 4, $X_5$ = three-level family history of prostate cancer (none, first-degree affected relative, other affected relative), $X_6$ = race in three categories (non-Hispanic White, non-Hispanic Black, other), $X_7$ = baseline cigarette smoking status (never, current, former) and baseline age. Note that $W$, $X_5$, $X_6$, and $X_7$ are not in the Cox model but may be predictive of continuous PSA.

**Missing at Random Phase-Two Data**

Covariate information can be missing for individuals in the phase-two sample. This would be the case if, for example, blood samples from individuals in phase-two have already been used or were lost. Suppose such covariates are missing at random conditional on $J^{(3)}$ strata that are based on phase-one or phase-two information and need not coincide with the $J$ phase-two strata. We can regard the set of individuals with complete covariate data as a *phase-three sample*. We let $w_{i,j}^{(3)} \equiv \frac{1}{\pi_{i,j}^{(3)}}$ be the phase-three design weight, where $\pi_{i,j}^{(3)}$ is the phase-three design sampling probability. The phase-three sampling probabilities may differ among the $J^{(3)}$ phase-three strata. The overall design weight of subject $i$ in stratum $j$ is $w_{i,j} = w_{i,j}^{(2)} \times w_{i,j}^{(3)}$. Usually, $\pi_{i,j}^{(3)}$ is unknown and $w_{i,j}^{(3)}$ must be estimated. Etiévant and Gail (12) provide details and take the estimation of the phase-three weights into account in the variance estimation.

In the PLCO data set we consider, there is no missing at random phase-two data. We nevertheless analyze case-cohorts where we pretend that individuals in the phase-two sample may have missing covariate information in Web Appendix 9.

**Installation and Implementation of CaseCohortCoxSurvival**

The unified variance estimation method in (12) is implemented in the R package CaseCohortCoxSurvival. The package can be installed directly from the CRAN Repository with the first two lines of R code in **Figure 1**. The package requires support from the external package survival. Specifically, the package contains the function `caseCohortCoxSurvival` to estimate Cox model parameters (log-RH, cumulative baseline hazard, PR) and their variance ($\hat{V}$ and $\hat{V}_{\text{robust}}$) from case-cohort data, as described in the help documentation (18). In the following Section, we describe how to use the function `caseCohortCoxSurvival` and provide some examples of code to analyze the PLCO data.

**How to Use the caseCohortCoxSurvival Function**

**Data Example**

The first few rows of the PLCO data set are displayed in **Table 3**. Note that $X_2$ is available on all members of the PLCO cohort, but for case-cohort analyses we pretend it is only available on cases (status = 1) and on non-cases who are in the subcohort (subcohort = 1). We estimate $\boldsymbol{\beta} = (\beta_1, \beta_2, \beta_3, \beta_4)'$ and the two PRs for covariate profiles $\boldsymbol{x} = (1, -0.7, 0, 1)'$ and $\boldsymbol{x} = (0, 0.7, 1, 0)'$ using the following sampling designs and methods of analysis: the unstratified case-cohort with design weights; the stratified case-cohort with design weights; the stratified case-cohort with calibrated weights. See the commands in Figure 1, which also produces variance estimates $\hat{V}$ and $\hat{V}_{\text{Robust}}$.

**Arguments**

We illustrate with the analysis of the stratified case-cohort with calibrated weights. The corresponding data and required arguments for `caseCohortCoxSurvival` are in **Table 4** and in the third example of **Figure 1**. The other examples in **Figure 1** and Web Appendices 1, 2, 3, 4 and 7 give corresponding arguments and examples of R script for analysis of the unstratified case-cohort with design weights, the unstratified case-cohort with calibrated weights, stratified case-cohort with design weights, and the full cohort.

In **Figure 1**, `data = cohort` specifies the data frame that has one row for each of the $n$ members of the cohort, `status = "status"` specifies the column name in `data` giving the case status, `subcohort = "subcohort"` specifies the column name in `data` giving the indicators of whether the person is in the subcohort, and `strata = "W"` specifies the column name in `data` with the stratum value used for the sampling of the subcohort. In **Figure 1**, the argument `time = c("entryage", "exitage")` specifies the column names in `data` giving the start age and end age, and implies that the analysis is on the age scale. The arguments `cox.phase1 = c("X1", "X3", "X4")` and `cox.phase2 = "X2"` specify the column names in `data` giving the covariates to be included in the Cox model and respectively from phase-one and phase-two. The argument `calibrated = TRUE` indicates that weight calibration will be performed and calibrated weights will be used. In that case, the user can either provide the auxiliary variables through the argument `aux.vars` or let the caseCohortCoxSurvival function build the auxiliary variables. The construction of the auxiliary variables relies on phase-one predictions of the phase-two covariates; the user can either provide predicted values of the phase-two covariates through the argument `predicted.cox.phase2` or provide phase-one predictors of the phase-two covariates through the arguments `predict = TRUE` and `predictors.cox.phase2`. The argument `aux.method` specifies the algorithm to construct the auxiliary variables; with `aux.method = "Breslow"`, only auxiliary variables for RH are used. The default is `aux.method = "Shin"`, which also includes the Shin auxiliary variables for PR. In **Figure 1**, we display the arguments for the stratified case-cohort analysis with calibrated weights and phase-one predictors of the phase-two covariates. The default `aux.vars = NULL` and `predicted.cox.phase2 = NULL` and argument `predictors.cox.phase2 = list(X2 = c("X5", "W", "X6", "X7", "entryage"))` imply that the prediction is not provided externally but is to be obtained by weighted regression in phase-two data of $X_2$ on $X_5, W, X_6, X_7$, and age at entry. Argument `aux.method` is not specified because the default is desired. To run the stratified case-cohort analysis with design weights instead of calibrated weights, the user should specify `calibrated = FALSE`, which is also the default (see Web Appendix 1). `Tau1 = 58` and `Tau2 = 68` define the age range over which cumulative baseline hazard and PR are to be predicted. The argument `x = pr1` specifies the covariate vector $(1, -0.7, 0, 1)$ for which PR is to be calculated. The argument `subcohort.strata.counts` can be used to specify the fixed number of individuals sampled in each stratum of the cohort ($m^{(j)}$). In **Figure 1**, this argument is omitted, leading to the default calculation using the counts in each stratum of the subcohort. The argument `weights.phase2` can be used to specify the column name in `data` giving phase-two design weights. The default is using the counts in each stratum of the subcohort and of the cohort to compute the stratum-specific phase-two design weights. Further details on these arguments, and on additional arguments, are provided in the help documentation of package CaseCohortCoxSurvival (18).

An example of outputs returned by function `caseCohortCoxSurvival` is given in Web Appendix 6.

**Results**

The results of unstratified/uncalibrated, stratified/uncalibrated, and stratified/calibrated PLCO case-cohort analyses are displayed in **Table 5**. For reference, we note that the full cohort analysis yielded estimates $\hat{\beta}_1 = 0.382, \hat{\beta}_2 = 1.239, \hat{\beta}_3 = 0.365, \hat{\beta}_4 = 0.537, \hat{\pi}(58,68;(0,0.7,1,0)) = 0.0406$ and $\hat{\pi}(58,68;(1,-0.7,0,1)) = 0.126$; see Web Appendix 8. Focusing on $\hat{V}$, the preferred variance estimate, we see that stratification improved efficiency compared to the unstratified analysis, both for RH and PR parameters. Weight calibration improved efficiency further for the log-RHs of covariates that were available on the entire cohort and for PRs. Calibration of the stratified design weights did not improve the efficiency of $\hat{\beta}_2$, because stratification of the subcohort is already based on a good predictor of $X_2$, $W$.

When using stratified uncalibrated design weights, $\hat{V}_{\text{Robust}} > \hat{V}$ for $\hat{\beta}_2$, but $\hat{V}_{\text{Robust}}$ agreed with $\hat{V}$ for $\hat{\beta}_1, \hat{\beta}_3$ and $\hat{\beta}_4$ because stratification is only based on a crude version of $X_2$. With uncalibrated design weights, $\hat{V}_{\text{Robust}} > \hat{V}$ for the pure risk estimates, not only for the stratified but also for the unstratified design. As anticipated from (12), $\hat{V}_{\text{Robust}}$ agreed with $\hat{V}$ for $\widehat{\boldsymbol{\beta}}$ with unstratified design weights, and $\hat{V}_{\text{Robust}}$ was very close to $\hat{V}$ for stratified calibrated weights for all parameters. Results of additional analyses are in Web Appendix 8.

**Other Features of the caseCohortCoxSurvival Function: Missing at Random Phase-Two Data and Computing PR for Various Covariate Profiles**

The unified variance estimation method in (12) also allows phase-two data to be missing at random for stratified or unstratified designs. This feature is available in the `caseCohortCoxSurvival` function from the R package [CaseCohortCoxSurvival](CaseCohortCoxSurvival). Details on the arguments required for such analyses are in Web Appendix 5, along with an example of R script in Web Appendix 9. We also display results of additional analyses of the PLCO data example where we consider stratified and unstratified case-cohorts with missing at random phase-two data in Web Appendix 10.

Several covariate profiles can be specified in the `x` argument of function `caseCohortCoxSurvival`. But by using the `casecohortcoxsurv` object returned by function `caseCohortCoxSurvival`, one can obtain PR parameter and variance estimates for a new covariate profile with the function `estimatePureRisk`, as in the first example of **Figure 1**.

## Discussion

We described how to use the R package [CaseCohortCoxSurvival](#) for inference on RH and PR with case-cohort sampling under the Cox model. This package implements the methods presented by Etiévant and Gail (12) and properly accounts for stratified or unstratified subcohort sampling with or without weight calibration to improve efficiency. It also handles phase-two covariates that are missing at random. Previous work (12) and examples in this paper illustrate that the widely used robust variance estimator (4,5) can be too large for RH in stratified designs and for PR in stratified and unstratified designs. We therefore recommend the variance calculations proposed by Etiévant and Gail (12) and produced by [CaseCohortCoxSurvival.](#)

We hope that this introduction to [CaseCohortCoxSurvival](#) will be useful to epidemiologists and statisticians who want to design or analyze a case-cohort study. The literature on the case-cohort design and its elaborations to include stratification and/or weight calibration is complex and technical, and the methods were not fully accessible through current software. Although there is some software online to estimate RH (7–9), code showing how to use survey software to accommodate weight calibration and stratification for RH seems no longer available online, and we could not find current R software for estimates of PR. Software like [CaseCohortCoxSurvival](#) makes available to the practitioner a convenient means for principled analyses tailored to various design and analysis contingencies. Thus, the software can help arrive at a valid analysis. The software can also promote better reporting of the design and analysis, because researchers need to specify these parameters as arguments to CaseCohortCoxSurvival. A review of publications using the case-cohort design indicates that important design and analysis features are often omitted (6). Finally, the software can promote better design choices, such as stratification and the gathering of ancillary information for weight calibration. For example, in contemplating a new study one could estimate desired parameters from pertinent simulated or real data with various stratification and weight calibration strategies for the case-cohort. By comparing the precision of estimates of key parameters across the various sampling and analysis strategies, computed with CaseCohortCoxSurvival, one can select a favorable sampling design and analysis approach. To obtain these several advantages, the user only needs to know how to supply the correct arguments to CaseCohortCoxSurvival and need not be concerned with theoretical and computational details. The algorithms in CaseCohortCoxSurvival are fast, because some of the computations take place in C subroutines. For example, the stratified calibrated analysis in **Table 5** took 15.82 seconds with R version 4.2.2 (2022-10-31) running on a Macbook Pro laptop (macOS Big Sur 10.16 with Process Quad-Core Intel Core i7 2.3 GHz). The associated memory management also allows one to analyze large cohorts.

In this article we assumed subcohort members were sampled *without replacement* from each stratum of the cohort, or from the whole cohort if there was no stratification. If sampling is *with replacement* (i.e. independent Bernoulli sampling), $\hat{V}_{\text{Robust}}$ is nearly equal to $\hat{V}$ and the output from CaseCohortCoxSurvival for $\hat{V}_{\text{Robust}}$ can be used (12).

## Acknowledgments

The authors thank Bill Wheeler at the Information Management Services, Inc.

## Author Affiliations

Lola Etiévant and Mitchell H. Gail: Biostatistics Branch, National Cancer Institute, Division of Cancer Epidemiology and Genetics, Rockville, Maryland, United States.

## Author Contributions

Conceptualization, L.E. and M.G.; writing- original draft preparation, L.E.; writing- review and editing, L.E. and M.G.; All authors have read and agreed to the published version of the manuscript.

## Grants and/or Financial Support

This work was supported by the Intramural Research Program of the Division of Cancer Epidemiology and Genetics, National Cancer Institute, National Institutes of Health, USA.

## Data-Availability Statement

The data underlying this article are maintained by the National Cancer Institute, Division of Cancer Epidemiology and Genetics and Division of Cancer Prevention, and are available to bona fide researchers upon submission and approval of a research proposal, and subsequent completion of a Data Transfer Agreement. Proposals can be submitted [here](here). The R package CaseCohortCoxSurvival is available on [CRAN](CRAN).

**Conflict of Interest Statement**

The authors declare no conflict of interest.

## Abbreviations

CRAN- the Comprehensive R Archive Network

RH- relative hazard

PLCO- Prostate, Lung, Colorectal and Ovarian Cancer Screening Trial

PR- pure risk

**Table 1.** The different case-cohort analyses covered by the R package CaseCohortCoxSurvival

| Unstratified subcohort | Stratified subcohort | Design weights | Calibrated weights | Missing data in phase-two |
|---|---|---|---|---|
| X |   | X |   |   |
|   | X | X |   |   |
| X |   |   | X |   |
|   | X |   | X |   |
| X |   |   |   | X |
|   | X |   |   | X |

**Table 2.** Stratified and unstratified sampling design for the PLCO analysis

| | Stratified subcohort | | | | Unstratified subcohort |
|---|---|---|---|---|---|
| | PSA $\leq$ 4 | PSA $>$ 4 | | | |
| Stratum $j$ | 1 | 2 | | | |
| $n_j$ | 30,882 | 2,665 | | $n$ | 33547 |
| $m_j$ | 3,177 | 2,380 | | $m$ | 4719 |
| Non-case design weight $w_j$ | 9.72 | 1.12 | | Non-case design weight $w$ | 7.11 |

**Figure 1.** Example of R script using the R package CaseCohortCoxSurvival to obtain parameter estimates and variance estimates log-RH and PR as in **Table 5**

```
install.packages("CaseCohortCoxSurvival")           # Install the R package
library("CaseCohortCoxSurvival")                    # Load the R package
load("PLCO.RData")                                  # Load the data set

# Covariate profiles for the PR ---------------------------------------
pr1 <- as.data.frame(cbind(X1 = 1, X2 = -0.7, X3 = 0, X4 = 1))
pr2 <- as.data.frame(cbind(X1 = 0, X2 = 0.7, X3 = 1, X4 = 0))

# Estimation using the unstratified case-cohort with design weights ----------
esti.unstrat.casecohort <- caseCohortCoxSurvival(data = cohort, status =
"status", subcohort = "unstrat.subcohort", time = c("entryage", "exitage"),
cox.phase1 = c("X1", "X3", "X4"), cox.phase2 = "X2", Tau1 = 58, Tau2 = 68, x =
pr1)

estimatePureRisk(esti.unstrat.casecohort, x = pr2)   # compute PR for another
covariate profile

# Estimation using the stratified case-cohort with design weights ------------
esti.casecohort <- caseCohortCoxSurvival(data = cohort, status = "status",
subcohort = "subcohort", strata = "W", time = c("entryage", "exitage"),
cox.phase1 = c("X1", "X3", "X4"), cox.phase2 = "X2", Tau1 = 58, Tau2 = 68, x =
rbind(pr1,pr2))

# Estimation using the stratified case-cohort with calibrated weights --------
esti.casecohort.calib <- caseCohortCoxSurvival(data = cohort, status =
"status", subcohort = "subcohort", strata = "W", time = c("entryage",
"exitage"), cox.phase1 = c("X1", "X3", "X4"), cox.phase2 = "X2", Tau1 = 58,
Tau2 = 68, calibrated = TRUE, predict = TRUE, predictors.cox.phase2 = list(X2
= c("X5", "W", "X6", "X7", "entryage")), x = rbind(pr1,pr2))
```

**Table 3.** First few rows of the PLCO data set

| | | $X_1$[a] | $X_2$[a,b] | $X_3$[a] | $X_4$[a] | $W$[a] | $X_5$[a] | $X_6$[a] | $X_7$[a] | entryage | exitage | status[c] | subcohort[d] |
|---|---|---|---|---|---|---|---|---|---|---|---|---|---|
| Individual $i$ | 1 | 0 | 0.45 | 0 | 1 | 0 | 2 | 1 | 0 | 67 | 80 | 0 | 0 |
| | 2 | 0 | 0.78 | 1 | 0 | 0 | 0 | 0 | 1 | 62 | 69 | 1 | 0 |
| | 3 | 0 | 1.78 | 1 | 0 | 1 | 0 | 0 | 2 | 64 | 72 | 0 | 1 |
| | 4 | 1 | -1.11 | 0 | 0 | 0 | 1 | 2 | 2 | 61 | 73 | 0 | 0 |
| | … | | | | | | | | | | | | |

[a] $X_1$ = indicator of first-degree family history of prostate cancer, $X_2$ = logarithm of continuous prostate specific antigen level in ng/mL, $X_3$ = indicator of being non-Hispanic White, $X_4$ = indicator of being non-Hispanic Black, $W$ = indicator of PSA > 4, $X_5$ = three-level family history of prostate cancer (none, first-degree affected relative, other affected relative), $X_6$ = race in three categories (non-Hispanic White, non-Hispanic Black, other), $X_7$ = baseline cigarette smoking status (never, current, former)

[b] To permit comparison with the full cohort analysis, we include values of $X_2$ for all cohort members, but $X_2$ would not be available in case-cohort (phase-two) studies, except for cases and for non-cases in the subcohort.

[d] This column indicates case status.

[d] This column indicates sampling into the subcohort.

**Table 4.** Details on the arguments to specify to the `caseCohortCoxSurvival` function to run the analysis of stratified case-cohort data with calibrated weights as in **Figure 1**

| | |
|---|---|
| `data` | Data frame containing the cohort and all variables needed for the analysis |
| `status` | Column name in `data` giving the case status for each individual in the cohort |
| `time` | Column name(s) in `data` giving the time to event for each individual in the case-cohort. One variable is required for a time-on-study time scale, two variables for age-scale, with the first variable as the start age and second as the end age. |
| `cox.phase1` | Column name(s) in `data` giving the Cox model covariates measured on the entire cohort |
| `cox.phase2` | Column name(s) in `data` giving the Cox model covariates measured only on phase-two individuals |
| `subcohort` | `NULL` or column name in `data` giving the indicators of membership in the subcohort. The indicators are 1 if the individual belongs to the subcohort and 0 otherwise. Some cases might be in the subcohort and others not. The default is `NULL`. If `NULL`, then a whole cohort analysis is performed. |
| `strata` | `NULL` or column name in `data` with the stratum value for each individual in the cohort. The number of strata used for the |
| | sampling of the subcohort equals the number of different stratum values. The default is NULL. |
| Subcohort.strata.counts | NULL or a named list giving the number of individuals sampled into the subcohort from each stratum of strata. The names in the list must be the strata values and the length of the list must be equal to the number of strata. The default is NULL. If NULL, the counts are estimated by the numbers of subcohort individuals in each stratum. |
| weights.phase2 | NULL or column name in data giving the phase-two design weights for each individual in the cohort. For a whole cohort analysis, weights are not used. The default is NULL. If NULL but subcohort is not NULL, the numbers of subcohort and cohort individuals in each stratum are used to estimate weights.phase2. |
| calibrated [a,b,c] | TRUE or FALSE to calibrate the weights. The default is FALSE. |
| predict [b,c] | TRUE or FALSE to predict the phase-two covariates using predictors.cox.phase2. The default is TRUE. |
| predicted.cox.phase2 [b,c] | NULL or a named list specifying the columns in data giving the predicted values of the phase-two covariates (cox.phase2) on the cohort. For example, if the phase-two covariates are X1 and X2, then the list is of the form list(X1=X1.pred, X2=X2.pred), where X1.pred and X2.pred are the names of |
| | the columns in `data` giving the predictions of `X1` and `X2` respectively. The default is `NULL`. |
| `predictors.cox.phase2` [b] | `NULL`, a vector or a list specifying the columns in `data` to use as predictor variables for obtaining the predicted values on the cohort for the phase-two covariates (`cox.phase2`). A named list allows for different phase-one variables to be used for the different phase-two covariates. For example, if the phase-two covariates are `X1` and `X2`, then the list is of the form `list(X1=c("X1.proxy"),X2=c("X1.proxy","X2.proxy"))`, where `X1.proxy` and `X2.proxy` are the names of columns in `data` giving phase-one predictors of `X1` and `X2`. The default is `NULL`. |
| `aux.vars` [b,c] | `NULL` or column name(s) in `data` giving the auxiliary variables for each individual in the cohort. The default is `NULL`. |
| `aux.method` | `"Breslow"`, or `"Shin"` to specify the algorithm to construct the auxiliary variables. `"Breslow"` only includes auxiliary variables for relative hazard, whereas `"Shin"`, the default, also has an auxiliary for pure risk. |
| `Tau1` | `NULL` or left bound of the time interval considered for the cumulative baseline hazard and pure risk. The default is `NULL`. If `NULL`, then the first event time is used. |
| Tau2 | NULL or right bound of the time interval considered for the cumulative baseline hazard and pure risk. The default is NULL. If NULL, then the last event time is used. |
| x | Data frame containing cox.phase1 and cox.phase2 variables for which pure risk is estimated. The default is NULL so that no pure risk estimates is computed. |

[a] If calibrated = FALSE, the stratified case-cohort data is analyzed with design weights. See also Web Appendix 1 and Web Appendix 2.

[b] Prediction of phase-two covariates is performed from predictors.cox.phase2 when calibrated = TRUE, predict = TRUE, aux.vars = NULL and predicted.cox.phase2 = NULL. See also Web Appendix 3.

[c] The predicted values of phase-two covariates predicted.cox.phase2 are used when calibrated = TRUE and aux.vars = NULL.

**Table 5.** Parameter and variance estimates of log-RH and PR with different sampling designs, methods of analysis and variance estimation methods in the PLCO data example

| | | Unstratified case cohort with design weights | | | Stratified case cohort with design weights | | | Stratified case cohort with calibrated weights | | |
|---|---|---|---|---|---|---|---|---|---|---|
| | | Parameter estimate | Variance estimate | | Parameter estimate | Variance estimate | | Parameter estimate | Variance estimate | |
| | | | $\hat{V}$ | $\hat{V}_{robust}$ | | $\hat{V}$ | $\hat{V}_{robust}$ | | $\hat{V}$ | $\hat{V}_{robust}$ |
| Log-relative hazard | $\beta_1$ | 0.353 | 0.00904 | 0.00904 | 0.317 | 0.00599 | 0.00599 | 0.339 | 0.00383 | 0.00383 |
| | $\beta_2$ | 1.231 | 0.00182 | 0.00182 | 1.238 | 0.00102 | 0.00105 | 1.235 | 0.00106 | 0.00106 |
| | $\beta_3$ | 0.225 | 0.0124 | 0.0124 | 0.43 | 0.0098 | 0.0098 | 0.394 | 0.00713 | 0.00713 |
| | $\beta_4$ | 0.426 | 0.0310 | 0.0310 | 0.561 | 0.0249 | 0.0249 | 0.583 | 0.0193 | 0.0193 |
| Pure risk in age interval (58,68] and for covariate profile $x$ | $x = (0, 0.7, 1, 0)$ | 0.0422 | 5.01E-05 | 5.04E-05 | 0.0369 | 2.46E-05 | 2.51E-05 | 0.0403 | 2.01E-05 | 2.01E-05 |
| | $x = (1, -0.7, 0, 1)$ | 0.130 | 1.95E-05 | 2.14E-05 | 0.127 | 1.49E-05 | 1.70E-05 | 0.125 | 1.30E-05 | 1.30E-05 |

# Web Appendices to CaseCohortCoxSurvival: an R Package for Case-Cohort Inference for Relative Hazard and Pure Risk under the Cox Model

Lola Etievant[1]*, Mitchell H. Gail[1]*

lola.etievant@nih.gov     ORCID 0000-0001-7562-3550

gailm@mail.nih.gov     ORCID 0000-0002-3919-3263

[1] National Cancer Institute, Division of Cancer Epidemiology and Genetics, Biostatistics Branch, 9609 Medical Center Drive, Rockville, MD 20850-9780.

* Corresponding authors

# Table of contents

# Web Appendix 1. Arguments to specify to the `caseCohortCoxSurvival` function to run the analysis of stratified case-cohort data with design weights

| | |
|---|---|
| `data` | Data frame containing the cohort and all variables needed for the analysis |
| `status` | Column name in `data` giving the case status for each individual in the cohort |
| `time` | Column name(s) in `data` giving the time to event for each individual in the case-cohort. One variable is required for a time-on-study time scale, two variables for age-scale, with the first variable as the start age and second as the end age. |
| `cox.phase1` | Column name(s) in `data` giving the Cox model covariates measured on the entire cohort |
| `cox.phase2` | Column name(s) in `data` giving the Cox model covariates measured only on phase-two individuals |
| `subcohort` | `NULL` or column name in `data` giving the indicators of membership in the subcohort. The indicators are 1 if the individual belongs to the subcohort and 0 otherwise. Some cases might be in the subcohort and others not. The default is `NULL`. If `NULL`, then a whole cohort analysis is performed. |
| `strata` | `NULL` or column name in `data` with the stratum value for each individual in the cohort. The number of strata used for the sampling of the subcohort equals the number of different stratum values. The default is `NULL`. |
| `Subcohort.strata.counts` | `NULL` or a named list giving the number of individuals sampled into the subcohort from each stratum of `strata`. The names in the list must be the strata values and the length of the list must be |
| | equal to the number of strata. The default is `NULL`. If `NULL`, the counts are estimated by the numbers of subcohort individuals in each stratum. |
| `weights.phase2` | `NULL` or column name in data giving the phase-two design weights for each individual in the cohort. For a whole cohort analysis, weights are not used. The default is `NULL`. If `NULL` but `subcohort` is not `NULL`, the numbers of subcohort and cohort individuals in each phase-two stratum are used to estimate `weights.phase2`. |
| `Tau1` | `NULL` or left bound of the time interval considered for the cumulative baseline hazard and pure risk. The default is `NULL`. If `NULL`, then the first event time is used. |
| `Tau2` | `NULL` or right bound of the time interval considered for the cumulative baseline hazard and pure risk. The default is `NULL`. If `NULL`, then the last event time is used. |
| `x` | Data frame containing `cox.phase1` and `cox.phase2` variables for which pure risk is estimated. The default is `NULL` so that no pure risk estimates is computed. |

# Web Appendix 2. Arguments to specify to the `caseCohortCoxSurvival` function to run the analysis of unstratified case-cohort data with design weights

| | |
|---|---|
| `data` | Data frame containing the cohort and all variables needed for the analysis |
| `status` | Column name in `data` giving the case status for each individual in the cohort |
| `time` | Column name(s) in `data` giving the time to event for each individual in the case-cohort. One variable is required for a time-on-study time scale, two variables for age-scale, with the first variable as the start age and second as the end age. |
| `Cox.phase1` | Column name(s) in `data` giving the Cox model covariates measured on the entire cohort |
| `cox.phase2` | Column name(s) in `data` giving the Cox model covariates measured only on phase-two individuals |
| `subcohort` | `NULL` or column name in `data` giving the indicators of membership in the subcohort. The indicators are 1 if the individual belongs to the subcohort and 0 otherwise. Some cases might be in the subcohort and others not. The default is `NULL`. If `NULL`, then a whole cohort analysis is performed. |
| `weights.phase2` | `NULL` or column name in data giving the phase-two design weights for each individual in the cohort. For a whole cohort analysis, weights are not used. The default is `NULL`. If `NULL` but `subcohort` is not `NULL`, the numbers of subcohort and cohort individuals are used to estimate `weights.phase2`. |
| Tau1 | NULL or left bound of the time interval considered for the cumulative baseline hazard and pure risk. The default is NULL. If NULL, then the first event time is used. |
| Tau2 | NULL or right bound of the time interval considered for the cumulative baseline hazard and pure risk. The default is NULL. If NULL, then the last event time is used. |
| x | Data frame containing `cox.phase1` and `cox.phase2` variables for which pure risk is estimated. The default is NULL so that no pure risk estimates is computed. |

# Web Appendix 3. Arguments to specify to the `caseCohortCoxSurvival` function to run the analysis of unstratified case-cohort data with calibrated weights

| | |
|---|---|
| `data` | Data frame containing the cohort and all variables needed for the analysis |
| `status` | Column name in `data` giving the case status for each individual in the cohort |
| `time` | Column name(s) in `data` giving the time to event for each individual in the case-cohort. One variable is required for a time-on-study time scale, two variables for age-scale, with the first variable as the start age and second as the end age. |
| `cox.phase1` | Column name(s) in `data` giving the Cox model covariates measured on the entire cohort |
| `cox.phase2` | Column name(s) in `data` giving the Cox model covariates measured only on phase-two individuals |
| `subcohort` | `NULL` or column name in `data` giving the indicators of membership in the subcohort. The indicators are 1 if the individual belongs to the subcohort and 0 otherwise. Some cases might be in the subcohort and others not. The default is `NULL`. If `NULL`, then a whole cohort analysis is performed. |
| `weights.phase2` | `NULL` or column name in data giving the phase-two design weights for each individual in the cohort. For a whole cohort analysis, weights are not used. The default is `NULL`. If `NULL` but `subcohort` is not `NULL`, the numbers of subcohort and cohort individuals are used to estimate `weights.phase2`. |
| `calibrated` [a,b,c] | `TRUE` or `FALSE` to calibrate the `weights`. The default is `FALSE`. |
| `predict` [b,c] | `TRUE` or `FALSE` to predict the phase-two covariates using `predictors.cox.phase2`. The default is `TRUE`. |
| predicted.cox.phase2 [b,c] | NULL or a named list specifying the columns in data giving the predicted values of the phase-two covariates (cox.phase2) on the cohort. For example, if the phase-two covariates are X1 and X2, then the list is of the form list(X1=X1.pred, X2=X2.pred), where X1.pred and X2.pred are the names of the columns in data giving the predictions of X1 and X2 respectively. The default is NULL. |
| predictors.cox.phase2 [b] | NULL, a vector or a list specifying the columns in data to use as predictor variables for obtaining the predicted values on the cohort for the phase-two covariates (cox.phase2). A named list allows for different phase-one variables to be used for the different phase-two covariates. For example, if the phase-two covariates are X1 and X2, then the list is of the form list(X1=c("X1.proxy"),X2=c("X1.proxy","X2.proxy")), where X1.proxy and X2.proxy are the names of columns in data giving phase-one predictors of X1 and X2. The default is NULL. |
| aux.vars [b,c] | NULL or column name(s) in data giving the auxiliary variables for each individual in the cohort. The default is NULL. |
| aux.method | "Breslow", or "Shin" to specify the algorithm to construct the auxiliary variables. "Breslow" only includes auxiliary variables for relative hazard, whereas "Shin", the default, also has an auxiliary for pure risk. |
| Tau1 | NULL or left bound of the time interval considered for the cumulative baseline hazard and pure risk. The default is NULL. If NULL, then the first event time is used. |
| Tau2 | NULL or right bound of the time interval considered for the cumulative baseline hazard and pure risk. The default is NULL. If NULL, then the last event time is used. |
| x | Data frame containing `cox.phase1` and `cox.phase2` variables for which pure risk is estimated. The default is `NULL` so that no pure risk estimates is computed. |

[a] If `calibrated = FALSE`, the case-cohort data is analyzed with design weights. See also Web Appendix 2.

[b] Prediction of phase-two covariates is performed from `predictors.cox.phase2` when `calibrated = TRUE, predict = TRUE, aux.vars = NULL` and `predicted.cox.phase2 = NULL`.

[c] The predicted values of phase-two covariates `predicted.cox.phase2` are used when `calibrated = TRUE` and `aux.vars = NULL`.

# Web Appendix 4. Arguments to specify to the `caseCohortCoxSurvival` function to run the analysis of cohort data

| | |
|---|---|
| `data` | Data frame containing the cohort and all variables needed for the analysis |
| `status` | Column name in `data` giving the case status for each individual in the cohort |
| `time` | Column name(s) in `data` giving the time to event for each individual in the case-cohort. One variable is required for a time-on-study time scale, two variables for age-scale, with the first variable as the start age and second as the end age. |
| `cox.phase1` | Column name(s) in `data` giving the Cox model covariates measured on the entire cohort |
| `Tau1` | `NULL` or left bound of the time interval considered for the cumulative baseline hazard and pure risk. The default is `NULL`. If `NULL`, then the first event time is used. |
| `Tau2` | `NULL` or right bound of the time interval considered for the cumulative baseline hazard and pure risk. The default is `NULL`. If `NULL`, then the last event time is used. |
| `x` | Data frame containing `cox.phase1` and `cox.phase2` variables for which pure risk is estimated. The default is `NULL` so that no pure risk estimates is computed. |

# Web Appendix 5. Arguments to specify to the `caseCohortCoxSurvival` function to run the analysis of stratified case-cohort data with missing at random phase-two data

| `data` | Data frame containing the cohort and all variables needed for the analysis |
|---|---|
| `status` | Column name in `data` giving the case status for each individual in the cohort |
| `time` | Column name(s) in `data` giving the time to event for each individual in the case-cohort. One variable is required for a time-on-study time scale, two variables for age-scale, with the first variable as the start age and second as the end age. |
| `cox.phase1` | Column name(s) in `data` giving the Cox model covariates measured on the entire cohort |
| `cox.phase2` | Column name(s) in `data` giving the Cox model covariates measured only on phase-two individuals |
| `subcohort` | `NULL` or column name in `data` giving the indicators of membership in the subcohort. The indicators are 1 if the individual belongs to the subcohort and 0 otherwise. Some cases might be in the subcohort and others not. The default is `NULL`. If `NULL`, then a whole cohort analysis is performed. |
| `strata` | `NULL` or column name in `data` with the stratum value for each individual in the cohort. The number of strata used for the sampling of the subcohort equals the number of different stratum values. The default is `NULL`. |
| `weights.phase2` | `NULL` or column name in data giving the phase-two design weights for each individual in the cohort. For a whole cohort analysis, |
| | weights are not used. The default is NULL. If NULL but subcohort is not NULL, the numbers of subcohort and cohort individuals in each stratum are used to estimate weights.phase2. |
| phase3 | NULL or column name in data giving the indicators of membership in the in the phase-three sample (i.e. having cox.phase2 measurements). The indicators are 1 if the individual belongs to the phase-three sample and 0 otherwise. All individuals in the phase-three sample must also belong to the phase-two sample. This option is not used if calibrated=TRUE. |
| strata.phase3 | NULL or column name in data giving the phase-three stratification for each individual in phase-two. The number of strata used for the third phase of sampling equals the number of different phase-three stratum values. |
| weights.phase3 | NULL or column name in data giving the phase-three design weights for each individual in phase-two. If NULL but phase3 is not NULL, then phase3 and subcohort are used to estimate weights.phase3. |
| weights.phase3.type | One of NULL, "design", "estimated", or "both" to specify whether the phase-three weights are known or to be estimated. The variance estimation differs for estimated and design weights. The default is "both". If set to "both", both variance estimates are computed. |
| Tau1 | NULL or left bound of the time interval considered for the cumulative baseline hazard and pure risk. The default is NULL. If NULL, then the first event time is used. |
| Tau2 | NULL or right bound of the time interval considered for the cumulative baseline hazard and pure risk. The default is NULL. If NULL, then the last event time is used. |
| x | Data frame containing `cox.phase1` and `cox.phase2` variables for which pure risk is estimated. The default is `NULL` so that no pure risk estimates is computed. |

# Web Appendix 6. Values returned by function `caseCohortCoxSurvival`

## a. When running a cohort analysis or case-cohort analysis without missing at random phase-two data

When `phase3 = NULL`, function `caseCohortCoxSurvival` returns a list with class `casecohortcoxsurv` that contains:

| | |
|---|---|
| `beta` | Log-relative hazard estimates |
| `beta.var` | Variance estimate $\hat{V}$ for `beta` |
| `beta.robustvar` | Robust variance estimate $\hat{V}_{\text{robust}}$ for `beta` |
| `Lambda0` | Cumulative baseline hazard estimate in interval (`Tau1`,`Tau2`] |
| `Lambda0.var` | Variance estimate $\hat{V}$ for `Lambda0` |
| `Lambda0.robustvar` | Robust variance estimate $\hat{V}_{\text{robust}}$ for `Lambda0` |
| `Pi.var` [a] | Matrix of pure risks and variance estimates $\hat{V}$ and $\hat{V}_{\text{robust}}$ for covariate profiles given in `x` and in interval (`Tau1`,`Tau2`] |

[a] If `x = NULL`, then `Pi.var` is not in the list.

## b. When running a cohort analysis or case-cohort analysis with missing at random phase-two data

When `phase3` is not `NULL`, function `caseCohortCoxSurvival` returns a list with class `casecohortcoxsurv` that contains:

| | |
|---|---|
| `beta` | Log-relative hazard estimates |
| `beta.var.estimated`[a] | Variance estimate $\hat{V}$ for `beta` when phase-three design weights are estimated. |
| `beta.robustvar.estimated`[a] | Robust variance estimate $\hat{V}_{robust}$ for `beta` when phase-three design weights are estimated. |
| `beta.var.design`[b] | Variance estimate $\hat{V}$ for `beta` when phase-three design weights are known. |
| `beta.robustvar.design`[b] | Robust variance estimate $\hat{V}_{robust}$ for `beta` when phase-three design weights are known. |
| `Lambda0` | Cumulative baseline hazard estimate in interval (`Tau1`,`Tau2`] |
| `Lambda0.var.estimated`[a] | Variance estimate $\hat{V}$ for `Lambda0` when phase-three design weights are estimated. |
| `Lambda0.robustvar.estimated`[a] | Robust variance estimate $\hat{V}_{robust}$ for `Lambda0` when phase-three design weights are estimated. |
| `Lambda0.var.design`[b] | Variance estimate $\hat{V}$ for `Lambda0` when phase-three design weights are known. |
| `Lambda0.robustvar.design`[b] | Robust variance estimate $\hat{V}_{robust}$ for `Lambda0` when phase-three design weights are known. |
| Pi.var.estimated [c] | Matrix of pure risks and variance estimates $\hat{V}$ and $\hat{V}_{robust}$ for covariate profiles given in `x` and in interval (`Tau1`,`Tau2`], when phase-three design weights are estimated. |
| Pi.var.design [d] | Matrix of pure risks and variance estimates $\hat{V}$ and $\hat{V}_{robust}$ for covariate profiles given in `x` and in interval (`Tau1`,`Tau2`], when phase-three design weights are known. |

[a] If `weights.phase3.type` = "design", then `beta.var.estimated`, `beta.robustvar.estimated`, `Lambda0.var.estimated` and `Lambda0.robustvar.estimated` are not in the list.

[a] If `weights.phase3.type` = "estimated", then `beta.var.design`, `beta.robustvar.design`, `Lambda0.var.design` and `Lambda0.robustvar.design` are not in the list.

[c] If `weights.phase3.type` = "design" or `x` = NULL, then `Pi.var.estimated` is not in the list.

[d] If `weights.phase3.type` = "estimated", or `x` = NULL, then `Pi.var.design` is not in the list.

# Web Appendix 7. R script to obtain log-relative hazard and pure risk estimates and variance estimates with the entire cohort and unstratified case-cohort with calibrated weights

## a. Rscript

```
library(CaseCohortCoxSurvival) # Load the R package ------------------------
load("PLCO.RData") # Load the data set ------------------------------------

# Covariate profiles for the pure risks-------------------------------------
pr1 <- as.data.frame(cbind(X1 = 1, X2 = -0.7, X3 = 0, X4 = 1))
pr2 <- as.data.frame(cbind(X1 = 0, X2 = 0.7, X3 = 1, X4 = 0))

# Estimation using the unstratified case-cohort with calibrated weights ------
esti.unstrat.casecohort.calib <- caseCohortCoxSurvival(data = cohort, status =
"status", subcohort = "unstrat.subcohort", time = c("entryage", "exitage"),
cox.phase1 = c("X1", "X3", "X4"), cox.phase2 = "X2", Tau1 = 58, Tau2 = 68,
calibrated = TRUE, predictors.cox.phase2 = list(X2 = c("X5", "W", "X6", "X7",
"entryage")), x = rbind(pr1,pr2))

# Estimation using the entire cohort ---------------------------------------
esti.cohort <- caseCohortCoxSurvival(data = cohort, status = "status", time =
c("entryage", "exitage"), cox.phase1 = c("X1", "X2", "X3", "X4"), Tau1 = 58,
Tau2 = 68, x = rbind(pr1,pr2))
```

Rscript to obtain log-relative hazard and pure risk estimates and variance estimates with the unstratified case-cohort with design weights, the stratified case-cohort with design weights and the stratified case-cohort with calibrated weights is given in Figure 1 in the Main Document.

## b. Example of outputs

```
> esti.unstrat.casecohort.calib$beta
X1          X3          X4          X2
0.4162564   0.2533165   0.4706341   1.2200619
```

```
> esti.unstrat.casecohort.calib$beta.var
X1                X3                X4                X2
0.003997168       0.007602208       0.020306929       0.001595534

> esti.unstrat.casecohort.calib$beta.robustvar
X1                X3                X4                X2
0.003997487       0.007601723       0.020307452       0.001596148

> esti.unstrat.casecohort.calib$Pi.var
      risk            variance          robust.variance
1     0.04506379      3.641817e-05      3.641660e-05
2     0.12631755      1.412048e-05      1.412098e-05

> esti.cohort$beta
X1            X2            X3            X4
0.3822328     1.2394166     0.3641509     0.5373042

> esti.cohort$beta.var
X1      X2      X3      X4
NA      NA      NA      NA

> esti.cohort$beta.robustvar
[1] 0.0027984109     0.0009925142     0.0063135879     0.0182014794

> esti.cohort$Pi.var
      risk            variance    robust.variance
1     0.04064131      NA          1.857796e-05
2     0.12628182      NA          1.059781e-05
```

# Web Appendix 8. Estimation with the entire cohort and with the unstratified case-cohort with calibrated weights in the PLCO data example

### a. Results

The estimation with the unstratified case-cohort with design weights, the stratified case-cohort with design weights and the stratified case-cohort with calibrated weights are displayed in Table 5 in the Main Document. Here, we also consider estimation with the unstratified case-cohort with calibrated weights, and as a point of reference, estimation with the full cohort ($n = 33,547$). We estimate the parameter variances by $\hat{V}$ and $\hat{V}_{\text{Robust}}$, and $\hat{V}_{\text{Robust}}$, respectively.

The Table in Web Appendix 8.b displays the parameter and variance estimates. With calibrated weights, $\hat{V}_{\text{Robust}}$ is very close to $\hat{V}$ for all parameters in the unstratified design too. In the unstratified design, using calibrated weights lead to an efficiency gain for all parameters, and the efficiency gain compared to estimation with the design weights is more important than in the stratified design. However, calibration in the stratified design leads to estimated variances that are smaller (and closer to the ones obtained with whole cohort) than in the unstratified design.

b. Table with parameter estimates and variance estimates of log-relative hazard and pure risk

|  |  | Unstratified case cohort with calibrated weights | | | Entire cohort | |
| --- | --- | --- | --- | --- | --- | --- |
|  |  | Parameter estimate | Variance estimate | | Parameter estimate | Variance estimate $\hat{V}_{robust}$ |
|  |  |  | $\hat{V}$ | $\hat{V}_{robust}$ |  |  |
| Log-relative hazard | $\beta_1$ | 0.416 | 0.004 | 0.004 | 0.382 | 0.0028 |
|  | $\beta_2$ | 1.22 | 0.0016 | 0.0016 | 1.239 | 0.00099 |
|  | $\beta_3$ | 0.253 | 0.0076 | 0.0076 | 0.365 | 0.00631 |
|  | $\beta_4$ | 0.471 | 0.0203 | 0.0203 | 0.537 | 0.0182 |
| Pure risk in age interval (58,68] and for covariate profile $x$ | $x = (0, 0.7, 1, 0)$ | 0.0451 | 3.64E-05 | 3.64E-05 | 0.0406 | 1.86E-05 |
|  | $x = (1, -0.7, 0, 1)$ | 0.126 | 1.41E-05 | 1.41E-05 | 0.126 | 1.06E-05 |

# Web Appendix 9. R script to obtain log-relative hazard and pure risk estimates and variance estimates from a case-cohort with missing at random phase-two data

a. R script

```r
install.packages("CaseCohortCoxSurvival")            # Install the R package
library("CaseCohortCoxSurvival")                     # Load the R package
load("PLCO.RData")                                   # Load the data set

# Covariate profiles for the pure risks ------------------------------------
pr1 <- as.data.frame(cbind(X1 = 1, X2 = -0.7, X3 = 0, X4 = 1))
pr2 <- as.data.frame(cbind(X1 = 0, X2 = 0.7, X3 = 1, X4 = 0))

# Estimation using the case-cohort with unstratified phase-two sampling,
# stratified phase-three sampling and known phase-three design weights ---------
esti.unstrat.casecohort.design <- caseCohortCoxSurvival(data = cohort, status
= "status", subcohort = "unstrat.subcohort", time = c("entryage", "exitage"),
cox.phase1 = c("X1", "X3", "X4"), cox.phase2 = "X2", strata.phase3 = "status",
phase3 = "unstrat.phase3", weights.phase3 = "unstrat.weights.phase3.true",
weights.phase3.type = "design", Tau1 = 58, Tau2 = 68, x = rbind(pr1,pr2))

# Estimation using the case-cohort with unstratified phase-two sampling,
# stratified phase-three sampling and estimated phase-three design weights -----
esti.unstrat.casecohort.estimated <- caseCohortCoxSurvival(data = cohort,
status = "status", subcohort = "unstrat.subcohort", time = c("entryage",
"exitage"), cox.phase1 = c("X1", "X3", "X4"), cox.phase2 = "X2", strata.phase3
= "status", phase3 = "unstrat.phase3", weights.phase3.type = "estimated", Tau1
= 58, Tau2 = 68, x = rbind(pr1,pr2))

# Estimation using the case-cohort with stratified phase-two and phase-three
# sampling and known phase-three design weights --------------------------------
esti.casecohort.design <- caseCohortCoxSurvival(data = cohort, status =
"status", subcohort = "subcohort", time = c("entryage", "exitage"), cox.phase1
= c("X1", "X3", "X4"), cox.phase2 = "X2", strata.phase3 = "status", phase3 =
"phase3", weights.phase3 = "weights.phase3.true", weights.phase3.type =
"design", Tau1 = 58, Tau2 = 68, x = rbind(pr1,pr2))
```

```r
# Estimation using the case-cohort with stratified phase-two and phase-three
sampling and estimated phase-three design weights --------------------------
esti.casecohort.estimated <- caseCohortCoxSurvival(data = cohort, status =
"status", subcohort = " subcohort", time = c("entryage", "exitage"), cox.phase1
= c("X1", "X3", "X4"), cox.phase2 = "X2", strata.phase3 = "status", phase3 = "
phase3", weights.phase3.type = "estimated", Tau1 = 58, Tau2 = 68, x =
rbind(pr1,pr2))
```

### b. Example of outputs

```
> esti.casecohort.design$beta
X1          X3          X4          X2
0.2924930   0.3962172   0.6626980   1.2675906

> esti.casecohort.design$beta.var.design
[1] 0.0063198453 0.0108076978 0.0201854224 0.0006077809

> esti.casecohort.design$beta.robustvar.design
[1] 0.0063188937 0.0108063543 0.0201828746 0.0006386685

> esti.casecohort.design$Pi.var.design
        risk           variance           robust.variance
1    0.03933265      2.574059e-05        2.627360e-05
2    0.12657098      1.519945e-05        1.727843e-05

> esti.casecohort.estimated$beta
X1          X3          X4          X2
0.2926153   0.3963804   0.6630475   1.2680004

> esti.casecohort.estimated$beta.var.estimated
X1               X3               X4               X2
0.0063270942     0.0108170006     0.0202054937     0.0006076356

> esti.casecohort.estimated$beta.robustvar.estimated
```

```
X1                X3                X4                X2
0.0063261410      0.0108156559      0.0202029430      0.0006385707

> esti.casecohort.estimated$Pi.var.estimated
      risk            variance          robust.variance
1     0.03924584      2.563823e-05      2.617041e-05
2     0.12633605      1.502300e-05      1.710186e-05
```

# Web Appendix 10. Estimation with the stratified and unstratified case-cohorts with missing at random phase-two data in the PLCO data example

## a. Data Example

As in the Main Document, the stratified phase-two sample is obtained by sampling $m^{(1)} = 3,177$ and $m^{(2)} = 2,380$ individuals from the $J = 2$ strata defined by $W$ in the entire cohort and then adding any other cases not in the subcohort. Similarly, the unstratified phase-two sample is obtained by sampling $m = 4719$ individuals from the entire cohort and adding the remaining cases. Now, we assume that individuals in the phase-two sample may have missing covariate information. More precisely, we assume that cases have a higher probability of missing covariate information than non-cases, for example because their stored blood samples have already been used. Therefore, we regard the case-cohort as a phase-three sample, obtained by further sampling individuals from the $J^{(3)} = 2$ strata defined by case status, with phase-three sampling probabilities $\boldsymbol{\pi}^{(3)} = (0.95, 0.9)$.

We estimate log-relative hazard $\boldsymbol{\beta} = (\beta_1, \beta_2, \beta_3, \beta_4)'$ and the pure risks for covariate profiles $\boldsymbol{x} = (1, -0.7, 0, 1)'$ and $\boldsymbol{x} = (0, 0.7, 1, 0)'$ using the following sampling designs and methods of analysis: the unstratified case-cohort with known phase-three design weights; the unstratified case-cohort with estimated phase-three design weights; the stratified case-cohort with known phase-three design weights; the stratified case-cohort with estimated phase-three design weights. Finally, for the variance estimation, we compare $\hat{V}_{\text{robust}}$ with $\hat{V}$. As a reminder, $\hat{V}$ properly accounts for the sampling features and for the variability from estimating of the phase-three weights (1). The corresponding arguments for `caseCohortCoxSurvival` are in Web Appendix 5 and the R code in Web Appendix 9.

## b. Results

The analysis results are displayed in the Tables in Web Appendix 10.c. For reference, we recall that the full cohort analysis yielded estimates $\hat{\beta}_1 = 0.382$, $\hat{\beta}_2 = 1.239$, $\hat{\beta}_3 = 0.365$, $\hat{\beta}_4 = 0.537$, $\hat{\pi}(58,68; (0,0.7,1,0)) = 0.0406$ and $\hat{\pi}(58,68; (1,-0.7,0,1)) = 0.126$; see Web Appendix 8. When using the case-cohort with a stratified second phase of sampling, $\hat{V}_{\text{Robust}} > \hat{V}$ for $\hat{\beta}_2$ and the pure risk estimates, but $\hat{V}_{\text{Robust}}$ agrees with $\hat{V}$ for $\hat{\beta}_1$, $\hat{\beta}_3$ and $\hat{\beta}_4$. When using the case-cohort with an unstratified second phase of sampling, $\hat{V}_{\text{Robust}} > \hat{V}$ for the pure risk estimates, and also for $\hat{\beta}_2$ when using estimated phase-three weights.

c. **Tables with parameter estimates and variance estimates of log-relative hazard and pure risk**

| | | Unstratified case cohort | | | | | |
|---|---|---|---|---|---|---|---|
| | | with known phase-three design weights | | | with estimated phase-three design weights | | |
| | | Parameter estimate | Variance estimate | | Parameter estimate | Variance estimate | |
| | | | $\hat{V}$ | $\hat{V}_{robust}$ | | $\hat{V}$ | $\hat{V}_{robust}$ |
| Log-relative hazard | $\beta_1$ | 0.481 | 0.00908 | 0.00908 | 0.481 | 0.00913 | 0.00913 |
| | $\beta_2$ | 1.256 | 0.00153 | 0.00153 | 1.258 | 0.00153 | 0.00154 |
| | $\beta_3$ | 0.572 | 0.0232 | 0.0232 | 0.573 | 0.0233 | 0.0233 |
| | $\beta_4$ | 0.692 | 0.0486 | 0.0486 | 0.694 | 0.0487 | 0.0487 |
| Pure risk in age interval (58,68] and for covariate profile $x$ | $x = (0,0.7,1,0)$ | 0.0414 | 4.69E-05 | 4.72E-05 | 0.0414 | 4.69E-05 | 4.71E-05 |
| | $x = (1,-0.7,0,1)$ | 0.126 | 1.79E-05 | 1.97E-05 | 0.126 | 1.78E-05 | 1.96E-05 |

| | | Stratified case cohort | | | | | |
|---|---|---|---|---|---|---|---|
| | | with known phase-three design weights | | | with estimated phase-three design weights | | |
| | | Parameter estimate | Variance estimate | | Parameter estimate | Variance estimate | |
| | | | $\hat{V}$ | $\hat{V}_{\text{robust}}$ | | $\hat{V}$ | $\hat{V}_{\text{robust}}$ |
| Log-relative hazard | $\beta_1$ | 0.292 | 0.00632 | 0.00632 | 0.293 | 0.00633 | 0.00633 |
| | $\beta_2$ | 1.268 | 0.00061 | 0.00064 | 1.268 | 0.00061 | 0.00064 |
| | $\beta_3$ | 0.396 | 0.0108 | 0.0108 | 0.396 | 0.0108 | 0.0108 |
| | $\beta_4$ | 0.663 | 0.0202 | 0.0202 | 0.663 | 0.0202 | 0.0202 |
| Pure risk in age interval (58,68] and for covariate profile $x$ | $x = (0, 0.7, 1, 0)$ | 0.0393 | 2.57E-05 | 2.63E-05 | 0.0392 | 2.56E-05 | 2.62E-05 |
| | $x = (1, -0.7, 0, 1)$ | 0.127 | 1.52E-05 | 1.73E-05 | 0.126 | 1.50E-05 | 1.71E-05 |

## Acknowledgments

The authors thank Bill Wheeler at the Information Management Services, Inc.

## Author Affiliations

Lola Etiévant and Mitchell H. Gail: Biostatistics Branch, National Cancer Institute, Division of Cancer Epidemiology and Genetics, Rockville, Maryland, United States.

## Author Contributions

Conceptualization, L.E. and M.G.; writing- original draft preparation, L.E.; writing- review and editing, L.E. and M.G.; All authors have read and agreed to the published version of the manuscript.

## Grants and/or Financial Support

This work was supported by the Intramural Research Program of the Division of Cancer Epidemiology and Genetics, National Cancer Institute, National Institutes of Health, USA.

## Data-Availability Statement

The data underlying this article are maintained by the National Cancer Institute, Division of Cancer Epidemiology and Genetics and Division of Cancer Prevention, and are available to bona fide researchers upon submission and approval of a research proposal, and subsequent completion of a Data Transfer Agreement. Proposals can be submitted [here](). The R package CaseCohortCoxSurvival is available on [CRAN]().

**Conflict of Interest Statement**

The authors declare no conflict of interest.